\documentclass[twocolumn,amsmath,amssymb,prb]{revtex4}
\usepackage{graphicx}
\usepackage{bm}

\begin{document}

\title{Resonance energy transfer near metal nanostructures mediated by surface plasmons}

\author{Vitaliy N. Pustovit$^{1,2}$ and Tigran V. Shahbazyan$^{3}$}

\affiliation{$^{1}$Centre de Recherche Paul Pascal, CNRS, Avenue A. Schweitzer, 33600 Pessac, France\\
$^{2}$Institute of Surface Chemistry, Kyiv 03164, Ukraine \\
$^{3}$Department of Physics, Jackson State University, Jackson, MS 39217, USA}


\begin{abstract}
We develop a unified theory of plasmon-assisted resonance energy transfer (RET) between  molecules  near a metal nanostructure that maintains energy balance between transfer, dissipation, and radiation. We show that in a wide range of parameters, including in the \textit{near field}, RET is dominated by \textit{plasmon-enhanced radiative transfer} (PERT) rather than by a nonradiative transfer mechanism. Our numerical calculations performed for molecules near the Ag nanoparticle indicate that RET magnitude is highly sensitive to molecules' positions.
\end{abstract}


\maketitle

\section{Introduction}
Resonance energy transfer (RET) between spatially separated molecules \cite{forster-ap48,dexter-jcp53} plays an important role in diverse phenomena across physics, chemistry and biology. Examples include photosynthesis, exciton transfer in molecular aggregates, energy exchange between proteins, \cite{lakowicz-book,andrews-book} and, more recently, between excitons in  quantum dots (QDs) \cite{klimov-prl02} and in QD-protein assemblies.\cite{willard-prl01} During past decade, significant advances were made in RET enhancement and control by placing molecules or QDs in microcavities \cite{hopmeier-prl99,andrew-science00,finlayson-cpl01} or near metal films and nanoparticles (NPs).\cite{leitner-cpl88,lakowicz-jf03,andrew-science04,lakowicz-jpcc07-1,lakowicz-jpcc07-2,krenn-nl08,rogach-apl08,yang-oe09,an-oe10} The coupling between molecular dipoles and surface plasmons (SP) in metal opens up new RET channels. The ability to control RET rates by adjusting dipoles' positions relative to metal surface is important in biomedical applications \cite{lakowicz-ab01} such as, e.g., SP  biosensors.\cite{kim-jacs05}

Near plasmonic nanostructure, RET from a donor to an acceptor is governed by the interplay between several processes. The energy of the excited donor   can either be radiated, dissipated,  or absorbed by the acceptor and each of these channels is affected by the nearby metal in its own way. In a closely related phenomenon, plasmon-enhanced fluorescence, the decay rates  in nonradiative and radiative channels depend differently on the distance between molecule and metal surface, $d$. The measured fluorescence\cite{feldmann-nl05,novotny-prl06,sandoghdar-prl06,halas-nl07} from molecules attached to a metal NP indeed shows that with decreasing $d$, SP enhancement is followed by quenching, in agreement with theory.\cite{nitzan-jcp81,ruppin-jcp82,pustovit-prl09,pustovit-prb10} A similar, albeit somewhat more complicated, scenario is expected when a donor \textit{and} an acceptor molecules are placed nearby a plasmonic nanostructure; i.e., the energy transfer from the  donor to the acceptor should be strongly affected by dissipation in metal and by plasmon-enhanced radiation. However, no RET theory including all  relevant energy flow channels has yet been available.  It is our goal   to provide such a theory here. 

To highlight the issue, recall the famous F\"{o}rster's formula for energy $W_{ad}^{F}$ transferred from donor to acceptor \cite{forster-ap48,dexter-jcp53,lakowicz-book,andrews-book}
\begin{align}
\label{fret-trad}
\frac{W_{ad}^{F}}{W_{d}}=\frac{9}{8\pi } \int \dfrac{d\omega }{k^4}f_d(\omega) 
\sigma_a(\omega) |D_{ad}^{0}|^{2},
\end{align}
where $W_{d}$ is the donor's radiated energy, $f_{d}(\omega)$ is its spectral function, $\sigma_{a}(\omega)$ is the acceptor's absorption crosssection,  $D_{ad}^{0}$ is the dipoles' electromagnetic coupling at distance $r_{ad}$  and $k$ is the wavevector of light. In the near field ($kr_{ad}\ll 1$),  we have $D_{ad}^{0}=q_{ad}/r_{ad}^{3}$ ($q_{ad}$ is the orientational factor) and RET  changes with distance as  $\left (r_{F}/r_{ad}\right )^{6}$, where $r_{F}$ is F\"{o}rster's radius.  In the far field ($kr_{ad}\gg 1$), RET is dominated by radiative coupling $|D_{ad}^{0}|\propto k^{2}/r_{ad}$ leading to weaker $r_{ad}^{-2}$ dependence.\cite{andrews-book,andrews-jcp92} Eq.~(\ref{fret-trad}) is derived from \textit{first-order} transition probability under the perturbation $D_{ad}^{0}$.

For molecules near a plasmonic nanostructure, Eq.~(\ref{fret-trad}) must be modified.  The standard model by Gersten and Nitzan \cite{nitzan-cpl84,nitzan-jcp85} and its extensions to planar and composite systems \cite{druger-jcp87,dung-pra02,stockman-njp08} incorporate SP in the transition's intermediate states, and thus Eq.~(\ref{fret-trad}) still holds albeit with new coupling $D_{ad}$ which now includes SP-mediated channels. However, this model accounts for neither  dissipation in metal  nor plasmon-enhanced radiation channels and, as a result, yields enormous (up to $10^{5}$) RET enhancement that contrasts  sharply with the much more modest ($\sim 10$)  increase \cite{lakowicz-jf03,andrew-science04,lakowicz-jpcc07-1,lakowicz-jpcc07-2,rogach-apl08,yang-oe09,an-oe10}  and even reduction \cite{leitner-cpl88,krenn-nl08} of measured RET rates.

\begin{figure}[bt]
  \begin{center}
  \includegraphics[width=0.99\columnwidth]{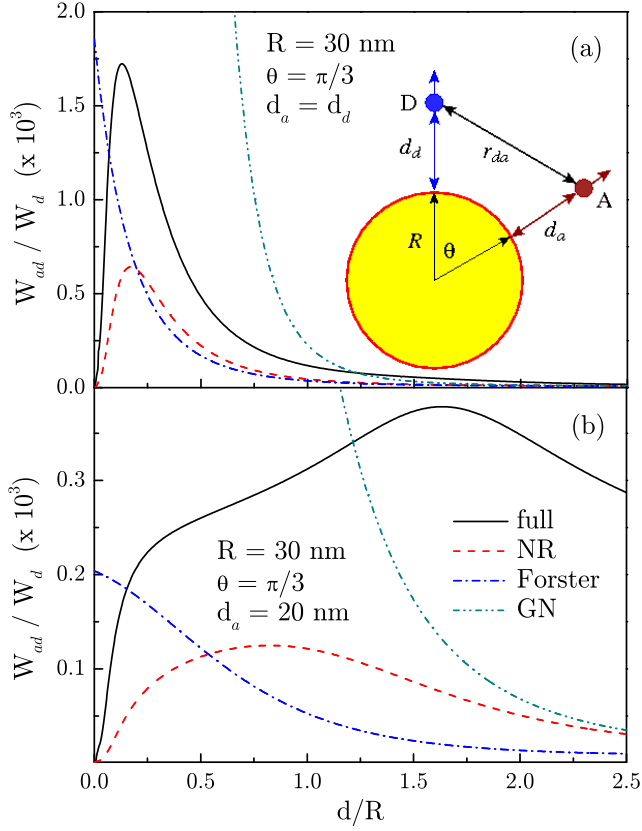}
  \end{center}
  \caption{\label{fig:dist_R30} (Color online) RET vs distance for $R=30$ nm Ag NP is shown at $\theta=\pi/3$  with (a) $d_{a}=d_{d}=d$ and (b) $d_{a}=20$ nm, $d_{d}=d$ using the full Eq.~(\ref{fret-new}), the nonradiative (NR) channel only, F\"{o}rster's transfer Eq.~(\ref{fret-trad}), and the Gersten-Nitzan (GN) model.\cite{nitzan-cpl84,nitzan-jcp85}}
  \end{figure}

Here we present a unified theory for RET near metal nanostructures based on the classical approach that accounts accurately for the full energy flow in the system. We show that Eq.~(1) is replaced with
\begin{align}
\label{fret-new}
\frac{W_{ad}}{W_{d}}=\frac{9}{8\pi } \int \dfrac{d\omega }{k^4}
\dfrac{\gamma_{d}^{r}}{\Gamma_{d}(\omega)}
\tilde{f}_{d}(\omega)\tilde{\sigma}_{a}(\omega) \left |\tilde{D}_{da}(\omega)\right |^{2},
\end{align}
where $\gamma_{d}^{r}$ is the donor's \textit{free space} radiative decay rate, $\Gamma_{d}$ is its \textit{full} decay rate, $\tilde{f}_{d}$ and $\tilde{\sigma}_{a}$ are the modified  spectral function and absorption cross section, respectively, and the coupling $\tilde{D}_{da}$ includes high-order SP-assisted transitions. For a low-yield donor, $\gamma_{d}^{r}$ should be replaced with the free space fluorescence rate $\gamma_{d}$. We also identify \textit{plasmon-enhanced radiative transfer} (PERT) as the dominant RET mechanism in a wide parameter range. In the \textit{far field}, we extract from Eq.~(\ref{fret-new}) a general formula for PERT from remote donors to an acceptor near the metal surface that extends  radiative RET theory \cite{andrews-book,andrews-jcp92} to plasmonic systems. In the \textit{near field}, our numerical calculations of RET near Ag NP (see inset in Fig.~\ref{fig:dist_R30}) show that PERT is the dominant mechanism here as well. Depending on system geometry, RET can  either be enhanced or reduced as compared to F\"{o}rster's transfer, consistent with experiment.\cite{leitner-cpl88,lakowicz-jf03,andrew-science04,lakowicz-jpcc07-1,lakowicz-jpcc07-2,krenn-nl08,rogach-apl08,yang-oe09,an-oe10}

%

\section{Theory of plasmon-assisted resonance energy transfer}
We consider a donor and an acceptor near the surface of a metal nanostructure which are represented by pointlike dipoles located  at ${\bf r}_{j}$ with induced moments ${\bf p}_j(\omega )=p_j(\omega) {\bf e}_{j}$ oriented along ${\bf e}_{j}$  ($j=a,d$). The dipoles are driven by the common electric field,
\begin{equation}
\label{moments}
{\bf p}_{j} (\omega ) =\alpha_{j} (\omega ) {\bf E}({\bf r}_{j}, \omega)+\delta_{jd}{\bf p}_{d}^{0}(\omega),
\end{equation}
where $\alpha_{j} (\omega )=\alpha'_{j} (\omega )+i\alpha''_{j} (\omega )$ is complex polarizability assumed  here isotropic, ${\bf p}_{d}^{0}(\omega ) = \alpha_{d} (\omega ) {\bf e}_{d} E_{0}$ is the donor's initial dipole moment with some constant $E_{0}$ depending on excitation, and $\delta_{jk}$ is Kroniker's symbol. The electric field $\bf E$ is, in turn, the solution of Maxwell's equation with dipole sources: \cite{novotny-book}
\begin{equation}
\label{field}
{\bf E}({\bf r},\omega ) =  \frac{4\pi \omega^2}{c^2} \sum_{j}  {\bf G}( {\bf r}, {\bf r}_{j}; \omega) \cdot {\bf p}_{j} (\omega ),
\end{equation}
where ${\bf G}({\bf r},{\bf r}';\omega)$ is Maxwell's equation Green's dyadic, satisfying ${\bm \nabla} \times {\bm\nabla} \times \hat{\bf G} - \epsilon({\bf r},\omega) (\omega/c)^{2}\hat{\bf G} = \hat{\bf I}$, and $\epsilon({\bf r},\omega)$ equals metal permittivity, $\epsilon(\omega)$, inside the metal region, and that of the outside medium, $\epsilon_{0}$, otherwise. The quantity of interest is energy absorbed by the acceptor in the unit frequency interval,
\begin{equation}
\label{dw} 
\frac{dW_{ad}}{d\omega}= -\frac{\omega}{\pi} {\rm Im} \left [{\bf p}^{*}_a(\omega) \cdot {\bf E}({\bf r}_a,\omega)\right ]=\frac{\omega \alpha''_{a}}{\pi}\left |\frac{p_{a}}{\alpha_{a}}\right |^{2},
\end{equation}
where we used ${\bf E}({\bf r}_a,\omega)={\bf p}_{a}(\omega)/\alpha_{a}(\omega)$ from Eq.~(\ref{moments}). A closed system for $p_{j}(\omega)$ is obtained by using Eq.~(\ref{field}) to eliminate the electric field from Eq.~(\ref{moments}),
\begin{equation}
\label{moments2}
 p_{j} (\omega ) + \alpha_{j} \sum_{k}  D_{jk} (\omega ) p_{k} (\omega ) 
= \delta_{jd} p_{d}^{0} (\omega ),
\end{equation}
where we introduce the frequency-dependent matrix 
\begin{equation}
\label{d-matrix}
D_{jk} (\omega ) = -\frac{4\pi \omega^2 }{c^2} 
{\bf e}_j  \cdot {\bf G}({\bf r}_{j},{\bf r}_{k};\omega) \cdot {\bf e}_k .
\end{equation}
Expressing $p_{a}$ from Eq.~(\ref{moments2}), we obtain
\begin{align}
\label{dw2}
\frac{dW_{ad}}{d\omega}
=\dfrac{\omega E_{0}^{2}}{\pi}\, \dfrac{\left |\tilde{\alpha}_{d}\right |^{2}\alpha''_a }{\left| 1+\alpha_a D_{aa} \right|^2}\left |\tilde{D}_{ad}\right |^{2},
\end{align}
where  $\tilde{D}_{ad}=D_{ad}\left [1-\tilde{\alpha}_{d}D_{da}\tilde{\alpha}_{a}D_{ad}\right ]^{-1}$ is donor-acceptor coupling that includes high-order transitions, and
\begin{equation}
\label{polar-dress}
\tilde \alpha_{j}(\omega)= \frac{\alpha_{j}(\omega)}{1+ D_{jj} (\omega)\alpha_{j}(\omega)}
\end{equation}
is  the molecule's \textit{dressed} polarizability satisfying the relation
\begin{equation}
\label{optical}
\tilde{\alpha}''_{j} + D''_{jj} |\tilde{\alpha}_{j}|^2 = \frac{\alpha''_{j}}{\left|1+ D_{jj} \alpha_{j}\right|^2},
\end{equation}
which expresses the \textit{energy balance} between total extinction described by $\tilde{\alpha}''_{j}$, external losses such as radiation and dissipation in metal encoded in  $D''_{jj}(\omega)$,  and absorption in the presence of environment (right-hand side). 

To gain more insight, recover first F\"{o}rster's RET from Eq. (\ref{dw2}). For  a \textit{high-yield} donor ($\alpha''_{d}=0$),  Eq.~(\ref{optical}) yields the optical theorem $\tilde{\alpha}''_{d0} =\frac{2}{3}k^{3}|\tilde{\alpha}_{d0}|^{2}$, where
\begin{equation}
\label{alpha0}
\tilde{\alpha}_{j0}=\dfrac{\alpha_{j}}{1-i \frac{2}{3} k^{3}\alpha_{j}}
\end{equation}
is polarizability in radiation field and we use free space expression for $D^{0}_{jj}= -i \frac{2}{3} k^{3}$. The near field coupling is 
\begin{equation}
D_{ad}^{0}=\left [{\bf e}_{a} \cdot {\bf e}_{d}-3({\bf e}_{a} \cdot \hat{\bf r}_{ad})({\bf e}_{d}\cdot \hat{\bf r}_{ad})\right ]/r_{ad}^{3}
\end{equation}
with $\hat{\bf r}={\bf r}/r$, while $\alpha_a D_{aa}^{0}\sim \alpha_ak^{3}$ is negligible. The radiated energy of an isolated donor can be derived in a similar manner as 
\begin{equation}
\label{Wd}
W_{d}=\frac{E_{0}^{2}}{\pi}\int d\omega \omega  \tilde {\alpha}''_{d0} (\omega).
\end{equation}
Using the optical theorem, Eq.~(\ref{dw2}) leads to  Eq.~(\ref{fret-trad}) with 
\begin{equation}
\label{sigma}
\sigma_{a}(\omega)=\frac{4\pi }{3}k\alpha''_{a}(\omega), ~~
 f_d(\omega)=\frac{\omega \tilde{\alpha}''_{d0}(\omega)}{\int d\omega \omega\tilde{\alpha}''_{d0} (\omega)},
\end{equation}
where the free space donor's spectral function $f_d(\omega)$ is integral-normalized to unity.

Turning to the general case, we note that for a high-yield donor, the energy balance relation Eq. (\ref{optical}) implies the \textit{optical theorem  in an absorptive environment},
\begin{equation}
\label{optical2}
\tilde{\alpha}''_{d} =-D''_{dd}|\tilde{\alpha}_{d}|^{2}=\frac{2}{3}k^{3}|\tilde{\alpha}_{d}|^{2}\frac{\Gamma_{d}}{\gamma_{d}^{r}},
\end{equation}
where $\Gamma_{j}=-\mu_{j}^{2}D''_{jj}$ is the molecule's \textit{full} decay rate \cite{novotny-book} and  $\gamma_{j}^{r}=\frac{2}{3}k^{3}\mu_{j}^{2}$ is its radiative decay rate ($\mu_{j}$ is the dipole matrix element). Using this relation and normalizing Eq.~(\ref{dw2}) to the radiated energy of an \textit{isolated} donor [Eq.~(\ref{Wd})], we obtain
\begin{equation}
\label{dw-final}
\frac{1}{W_{d}}\dfrac{dW_{ad}}{d\omega}
=\frac{9}{8 \pi k^4} \dfrac{\gamma_{d}^{r}}{\Gamma_{d}(\omega)}\tilde{f}_d(\omega) \tilde{\sigma}_{a}(\omega)\left |\tilde{D}_{da}\right |^{2},
\end{equation}
which leads to Eq.~(\ref{fret-new}) after frequency integration. Here
\begin{equation}
\label{sigma-modified}
\bar{\sigma}_{a}=\dfrac{4\pi k}{3}\dfrac{\alpha''_{a}}{\left | 1+\alpha_{a} D_{aa} \right|^{2}},
~~~ 
\tilde{f}_d(\omega)=\dfrac{\omega \tilde{\alpha}''_{d}(\omega)}{\int d\omega \omega \tilde{\alpha}''_{d0}(\omega)}
\end{equation}
are the acceptor's absorption cross section and the donor's spectral function modified by the environment [compare to Eq.~(\ref{sigma})]. Note that, in the presence of metal,  $\tilde{f}_d(\omega)$  is no longer integral-normalized to unity. 

Equation (\ref{dw-final}) includes \textit{all} relevant energy flow channels in the system.   Interactions of the molecules with the metal alter the positions and shapes of the optical bands. While the coupling $D_{ad}$ is enhanced due to plasmon-mediated channels, the factor $\gamma_{d}^{r}/\Gamma_{d}$ accounts for RET \textit{quenching} due to the donor's energy transfer to the metal followed by dissipation and radiation. The absence of this factor leads to spuriously large RET.\cite{nitzan-cpl84,nitzan-jcp85,druger-jcp87,dung-pra02,stockman-njp08} Note that Eq.~(\ref{fret-new}) was obtained for a high-yield donor with no assumptions on molecules' emission or absorption spectral bands, which are usually broad and asymmetric due to vibrational and rotational modes. Rigorous treatment of molecules' internal relaxation processes would require fully quantum-mechanical consideration which is beyond our scope. However, if we assume Lorenzian lineshape for the donor's effective polarizability $\tilde{\alpha}_{d}(\omega)$, which is a reasonable approximation in most cases, then it is easy to show that Eq.~(\ref{fret-new}) is valid for low-yield donors as well upon replacing $\gamma_{d}^{r}$ with the free space fluorescence rate $\gamma_{d}$.

To highlight the role of PERT in the \textit{far field} RET, consider energy transfer from \textit{remote} donors to an acceptor located \textit{near} the metal surface. In this case, the donor's decay rate and spectral function are unaffected by metal and RET is dominated by the following process: A donor first \textit{radiatively} excites SP in the metal which then \textit{nonradiatively} transfers its energy to the acceptor. The coupling $D_{ad}$ can be derived from Dyson's equation for Green's dyadic, 
\begin{equation}
{\bf G}({\bf r},{\bf r}')={\bf G}^{0}({\bf r},{\bf r}') +k^{2}\bar{\epsilon}\int dV_{m} {\bf G}^{0}({\bf r},{\bf r}_{m})\cdot {\bf G}({\bf r}_{m},{\bf r}'),
\end{equation}
where integration is restricted to  metal region and $\bar{\epsilon}(\omega)=\epsilon(\omega)/\epsilon_{0}-1$. For remote donors, using the far field limit ($kr\gg 1$ and $kr'\ll 1$) of the free Green's dyadic,\cite{novotny-book} ${\bf G}^{0}({\bf r},{\bf r}')=\frac{e^{ikr}}{4\pi r}(\delta_{\mu\nu}-\hat{\bf r}_{\mu}\hat{\bf r}_{\nu})$,  and averaging out over donors angular positions and their dipoles' orientations, we obtain PERT \textit{per donor}
\begin{align}
\label{fret-rad}
\frac{W_{ad}^{r}}{W_{d}}\approx \frac{1}{4\pi r_{ad}^{2}} \int d\omega f_d(\omega) 
\bar{\sigma}_{a}(\omega)A(\omega),
\end{align}
where 
\begin{equation}
A=\left |{\bf e}_{a}+k^{2}\bar{\epsilon}\int dV_{m}{\bf G}({\bf r}_{m},{\bf r}_{a})\cdot {\bf e}_{a}\right |^{2}
\end{equation}
is the SP enhancement factor for a metal nanostructure of general shape. If the acceptor is located at distance $r_{a}$ from the center of a spherical NP, we get $A=A^{\perp}\cos^{2}\phi+A^{\parallel}\sin^{2}\phi$, where  
\begin{equation}
A^{\perp}=\left |1+2\frac{\alpha_{1}}{r_{a}^{3}}\right |^{2},
~~
A^{\parallel}=\left |1-\frac{\alpha_{1}}{r_{a}^{3}}\right |^{2}
\end{equation}
%
are enhancement factors for normal and parallel dipole orientations,\cite{nitzan-jcp81}  $\alpha_{1}(\omega)$ is the NP dipole polarizability, and $\cos\phi=\hat{\bf r}_{a}\cdot {\bf e}_{a}$. Eq.~(\ref{fret-rad}) extends the \textit{far field} radiative RET theory \cite{andrews-book,andrews-jcp92} to plasmonic systems. In fact, the PERT mechanism can dominate RET even in the \textit{near field}, as our numerical calculations below demonstrate.

\section{Numerical results for near field energy transfer}
\begin{figure}[tb]
  \begin{center}
  \includegraphics[width=0.99\columnwidth]{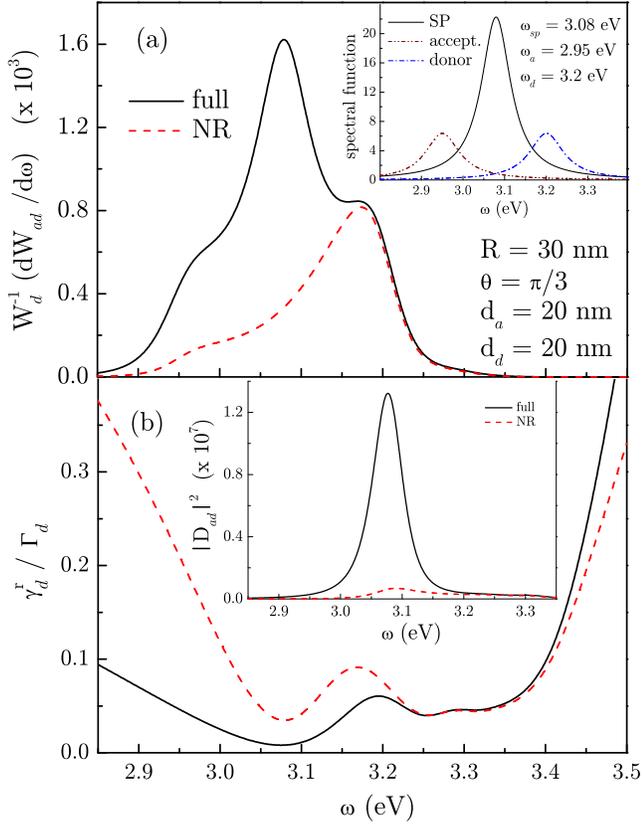}
  \end{center}
  \caption{\label{fig:sp_R30}  (Color online) (a) Spectral function Eq.~(\ref{dw-final})  and molecules' optical bands relative to SP band  $\alpha_{1}/R^{3}$ (inset)  are shown together with (b) quenching factor $\gamma_{d}^{r}/\Gamma_{d}$ and coupling $|D_{ad}|^{2}$ (inset) using full and nonradiative (NR) models.}
  \end{figure}

As an example, consider a donor and acceptor near spherical Ag NP in water with normal dipole orientations (see Fig. 1).  The near field matrix $D_{jk}$ is readily obtained from the Mie's theory Green's dyadic \cite{ruppin-jcp82} as  $D_{jk}=D_{jk}^{0}+D_{jk}^{r}+D_{jk}^{nr}$, where \cite{pustovit-prb10}
\begin{align}
\label{d-matrix-perp}
D_{jk}^{r}=
 &
 -i\frac{2}{3}k^{3}\biggl[1+ 2 \alpha_1 \biggl (\frac{1} {r_{j}^3}+ \frac{1} {r_{k}^3} \biggr ) 
+ \frac{4 |\alpha_1|^2}{r_{j}^3 r_{k}^3}\biggr ](\hat{\bf r}_{j}\cdot \hat{\bf r}_{k}),
\nonumber\\
D_{jk}^{nr} =
&
-\sum_{l}\frac{\alpha_{l} (l+1)^2  } {r_{j}^{l+2} r_{k}^{l+2}}P_{l} (\hat{\bf r}_{j}\cdot \hat{\bf r}_{k})
\end{align}
are NP-induced radiative and nonradiative terms,  $\alpha_{l}=R^{2l+1}\frac{l(\epsilon -\epsilon_0)}{l\epsilon +(l+1)\epsilon_0}$ is NP polarizability, $P_{l}(x)$ is a Legendre polynomial, $D_{ad}^{0}=( 1 + \sin^2 \theta/2)/r_{ad}^{3}$, $\hat{\bf r}_{a}\cdot \hat{\bf r}_{d}=\cos\theta $, and angular momenta  up to $l = 50$ are included. Full decay rates are  $\Gamma_{j}=-\left (3/2k^{3}\right )\gamma_{j}^{r}D''_{jj}$. We consider, for simplicity, a high-yield donor with a broad emission band due to the vibrational modes. Molecules optical bands are Lorentzians of width 0.05 eV centered at 2.95 eV and 3.2 eV with maximal overlap at about SP energy of 3.08 eV [see inset in Fig.~\ref{fig:sp_R30}(a)], $\sigma_{a}(\omega)$ was normalized to its total $\int d\omega\sigma_{a}(\omega)$, and modified $\bar{\sigma}_{a}$, $\tilde{f}_{d}$, and $\tilde{D}_{ad}$ were found using Eq.~(\ref{d-matrix-perp}). 
%
%
%

In Fig.~\ref{fig:dist_R30}, we plot $W_{ad}$ vs the molecule's distance $d$ from the $R=30$ nm NP surface at $\theta=\pi/3$ with equal $d_{a}=d_{d}=d$ and with changing $d_{d}=d$ at fixed $d_{a}$. Three  models---the full Eq.~(\ref{fret-new}), its nonradiative part only, and the Gersten-Nitzan model \cite{nitzan-cpl84,nitzan-jcp85}---are compared to F\"{o}rster's transfer Eq.~(\ref{fret-trad}).  For $d_{d}=d_{a}$, $W_{ad}$ is about three times larger than $W_{ad}^{F}$ and rapidly decays with $d$, while for $d/R\ll 1$ it is quenched by metal. There is no enhancement if only the nonradiative channel is included in Eq.~(\ref{fret-new}). In contrast, the Gersten-Nitzan model yields much greater enhancement (up to $10^{5}$) for $d/R \ll 1$ since it includes no quenching effects. However, at fixed $d_{a}$ and $d_{d}/R \gtrsim 1$, the full $W_{ad}$ is the largest one [see Fig.~\ref{fig:dist_R30}(b)] due to the dominant role of the PERT mechanism, as discussed above.

%
  \begin{figure}[tb]
 \begin{center}
  \includegraphics[width=0.99\columnwidth]{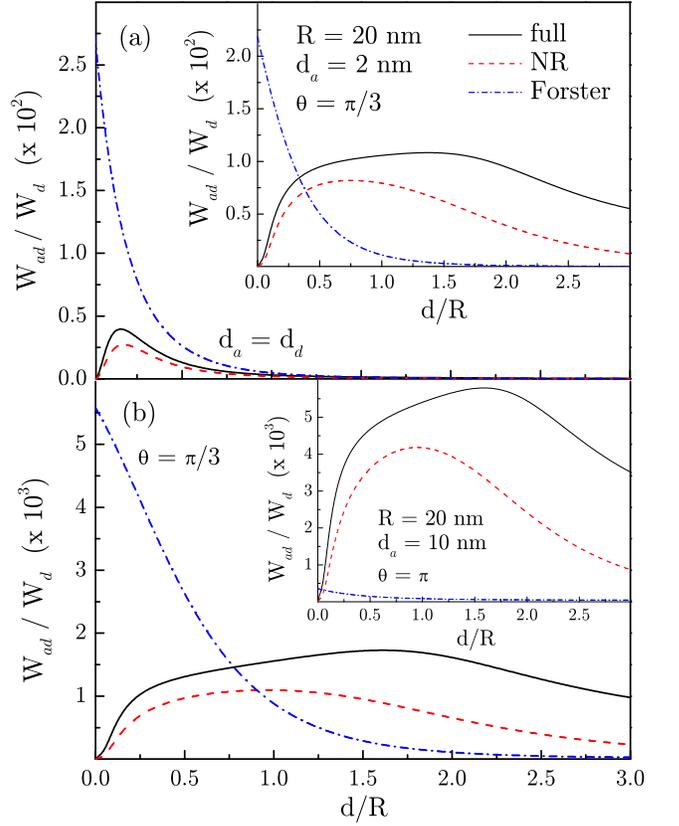}
  \end{center}
  \caption{\label{fig:dist_R20} (Color online) RET vs distance for $R=20$ nm Ag NP is shown (a) at $\theta=\pi/3$ with $d_{a}=d_{d}$ and $d_{a}=2$ nm (inset) and (b) with $d_{a}=10$ nm at $\theta = \pi/3$ and $\theta=\pi$ (inset) using full, nonradiative (NR), and F\"{o}rster models.}
  \end{figure}

The interplay of different RET contributions is shown in Fig.~\ref{fig:sp_R30} featuring spectral density Eq.~(\ref{dw-final}) together with quenching factor $\gamma_{d}^{r}/\Gamma_{d}$ and coupling $\left |D_{ad}\right |^{2}$ at fixed $d$. $dW_{ad}/d\omega$ has a sharp SP peak which disappears if only the nonradiative channel is included [see Fig.~\ref{fig:sp_R30}(a)].  PERT channel \textit{reduces} $\gamma_{d}^{r}/\Gamma_{d}$ due to SP-enhanced radiation but it strongly \textit{enhances} $D_{ad}$ [see Fig.~\ref{fig:sp_R30}(b)], the net result being RET increase, while in the nonradiative channel the enhancement and quenching effects nearly cancel out. Weak high-frequency oscillations are due to high-$l$ SPs.

The  relative rates of SP-assisted RET and F\"{o}rster's transfer are highly sensitive to the system's geometry. RET is quenched if both molecules are close to the NP surface [see Fig.~\ref{fig:dist_R20}(a)] but it becomes enhanced if donor-NP distance increases (inset). For  $\theta=\pi/3$ RET is enhanced if $d_{d}\gtrsim R$ [see Fig.~\ref{fig:dist_R20}(b)], but for $\theta=\pi$ it is strongly enhanced for nearly all $d$ (inset). In fact, NP acts as a \textit{hub} that couples equally well nearby and remote molecules with different $\theta$ while F\"{o}rster's transfer drops for large $r_{ad}$. For smaller NP sizes, the role of PERT becomes less pronounced yet remains dominant for larger donor-NP distances. 

\section{Conclusions}
In summary, a theory of resonance energy transfer between energy donors and acceptors near a plasmonic structure is presented which maintains a correct energy balance between transfer, dissipation, and radiation that is essential for interpretation of experimental data. The plasmon-enhanced radiative transfer is shown to be the dominant mechanism in a wide parameter range. This work was supported by the NSF under Grants No. DMR-0906945 and No. HRD-0833178, and by the EPSCOR program.

{}

\end{document}